\documentclass[a4paper]{article}
\usepackage[a4paper,top=3cm,bottom=4cm,left=2cm,right=2cm]{geometry}
\usepackage[utf8]{inputenc}

\usepackage{amsmath}
\usepackage{amssymb}
\usepackage{graphicx}
\usepackage{hyperref}
\hypersetup{
  colorlinks   = true, 
  urlcolor     = blue, 
  linkcolor    = blue, 
  citecolor   =  blue 
}
\usepackage{lineno}

\newcommand{\alglong}{Term Analysis with Graph Deep Q-Network}
\newcommand{\algshort}{TAG-DQN}

\title{Accelerating Atomic Fine Structure Determination \\ with Graph Reinforcement Learning}
\author{Milan Ding\thanks{Corresponding author: milan.ding15@imperial.ac.uk} \\
Department of Physics, Imperial College London, \\
Prince Consort Road, London, SW7 2AZ, United Kingdom 
\and
Victor-Alexandru Darvariu \\
Oxford Robotics Institute, University of Oxford, \\
17 Parks Road, Oxford, OX1 3PJ, United Kingdom
\and
Alexander N. Ryabtsev \\
Institute of Spectroscopy, Russian Academy of Sciences, \\
Troitsk, Moscow 108840, Russia
\and
Nick Hawes \\
Oxford Robotics Institute, University of Oxford, \\
17 Parks Road, Oxford, OX1 3PJ, United Kingdom
\and
Juliet C. Pickering \\
Department of Physics, Imperial College London, \\
Prince Consort Road, London, SW7 2AZ, United Kingdom 
}
\begin{document}
\maketitle 
\section*{Abstract}
Atomic data determined by analysis of observed atomic spectra are essential for plasma diagnostics. For each low-ionisation open d- and f-subshell atomic species, around $10^3$ fine structure level energies can be determined through years of analysis of $10^4$ observable spectral lines. We propose the automation of this task by casting the analysis procedure as a Markov decision process and solving it by graph reinforcement learning using reward functions learned on historical human decisions. In our evaluations on existing spectral line lists and theoretical calculations for Co II and Nd II-III, hundreds of level energies were computed within hours, agreeing with published values in 95\% of cases for Co~II and 54-87\% for Nd~II-III. As the current efficiency in atomic fine structure determination struggles to meet growing atomic data demands from astronomy and fusion science, our new artificial intelligence approach sets the stage for closing this gap.


\clearpage

\section{Introduction}
The determination of atomic fine structure of low-ionisation open d- and f-subshell elements involves extracting energies and total electron angular momenta $J$ of energy levels underlying observed atomic spectra. This process, commonly referred to as \textit{term analysis}~\cite{johansson1996term}, also assigns term symbols to levels. It is a sequential, complex decision-making task requiring both atomic spectroscopy expertise and extensive human labour. The resulting level energies and transition wavenumbers are fundamental data~\cite{kramida2024nist} underpinning plasma diagnostics within lighting and metal industries, magnetic confinement fusion research~\cite{muller2015}, and astrophysics~\cite{atkins2013, heiter2021}, including renewed interest in f-subshell species in neutron star merger observations that are transforming our understanding of the origin of heavy elements~\cite{cowan2021}. However, most levels and transitions for the heavier elements remain unknown \cite{kramida2024nist}, and advanced \textit{ab initio} calculations are accurate only to a few percent~\cite{cowan1981, gaigalas2019, kramida2014}, insufficient for applications requiring higher spectral resolutions. Term analyses are therefore vital, offering orders-of-magnitude higher accuracies and reliable theoretical model constraints.

Term analyses commonly involve Fourier transform (FT) and grating spectroscopy of plasmas under resolving powers up to $10^6$ and $10^5$, respectively, and dynamic ranges up to $10^4$ across the infrared, visible, and UV ranges~\cite{concepcion2023, reader2001, tchang-brillet2002}. A transition between an upper level with energy $E_u$ and a lower level with $E_l$ is observed as a spectral line at wavenumber $\sigma$ equal to the energy difference, $\sigma = E_u - E_l$. Several $10^4$ spectral lines are observable for each open d- or f-subshell element. The primary challenge is determining level energies from an immense number of energy differences, guided by less accurate theoretical predictions~\cite{johansson1996term, azarov1991, azarov1993}. Existing tools support spectral line wavenumber and intensity extraction~\cite{engstrom1998, nave2015, ding2025}, visualisation for manual level energy determination~\cite{azarov2018}, and level energy optimisation~\cite{kramida2011}. While spectrum measurement and theoretical calculations for one species can be completed in weeks, the subsequent analyses still require months to years and remain a major bottleneck. Since the advent of FT spectroscopic term analyses in the 1970s~\cite{connes1970spectroscopie}, its scope has been mainly limited to the iron-group ($23\leq Z\leq 28$) elements~\cite{kramida2024nist}.

Early research demonstrated the potential of pattern-recognition methods for partial term analysis procedures \cite{peterson1978spectrala, peterson1978spectralb}. Here, we develop new artificial intelligence (AI) techniques to automate term analyses by casting the problem as a Markov decision process (MDP) involving graphs, where an agent determines unknown levels (nodes) and lines (edges) over discrete time steps via reinforcement learning (RL). 
The agent learns to choose valid actions that maximise a reward function trained on human preferences from past analyses. 
We propose a variant of the Deep Q-network (DQN) algorithm~\cite{mnih2015} called \textit{\alglong}~(\algshort), which belongs to the graph RL~\cite{darvariu2024grl} class of methods for tackling combinatorial decision-making problems over graphs. Key to achieving scalability is the adoption of techniques that have proven successful in this broader literature including action space decompositions, restraining valid actions via domain knowledge, and using graph neural networks (GNNs) as a learning representation~\cite{scarselli2009}.

We emulated early to intermediate stages of published term analyses of Co~II \cite{pickering1998}, Nd~III \cite{ding2024}, and Nd~II \cite{blaise1984} as initial MDP states for evaluation case studies, using existing FT spectral line lists \cite{pickering1998, ding2024} and theoretical calculations achievable given the initial state known levels. Correct level energy determination rates ranged from 54\% with \textit{ab initio} atomic structure calculations to 95\% with semi-empirical methods. For the first time, up to $10^2$ tentative fine structure level energies can be computed in hours, alleviating months of human labour needed to achieve similar results. These results were enabled by \algshort~learning from rapid attempts at level energy determinations and resolving ambiguities by reaching MDP states (level systems) most consistent within experimental and theoretical uncertainties. Performance also compared favourably with baseline search algorithms. We present this work as the conception of a revolutionary tool for term analyses, enabling rapid developments in fundamental atomic data in years that would otherwise take decades, thereby accelerating progress in atomic physics, astronomy, and fusion technology.
\section{Results}

\begin{figure}
    \centering
    \includegraphics[width=\linewidth]{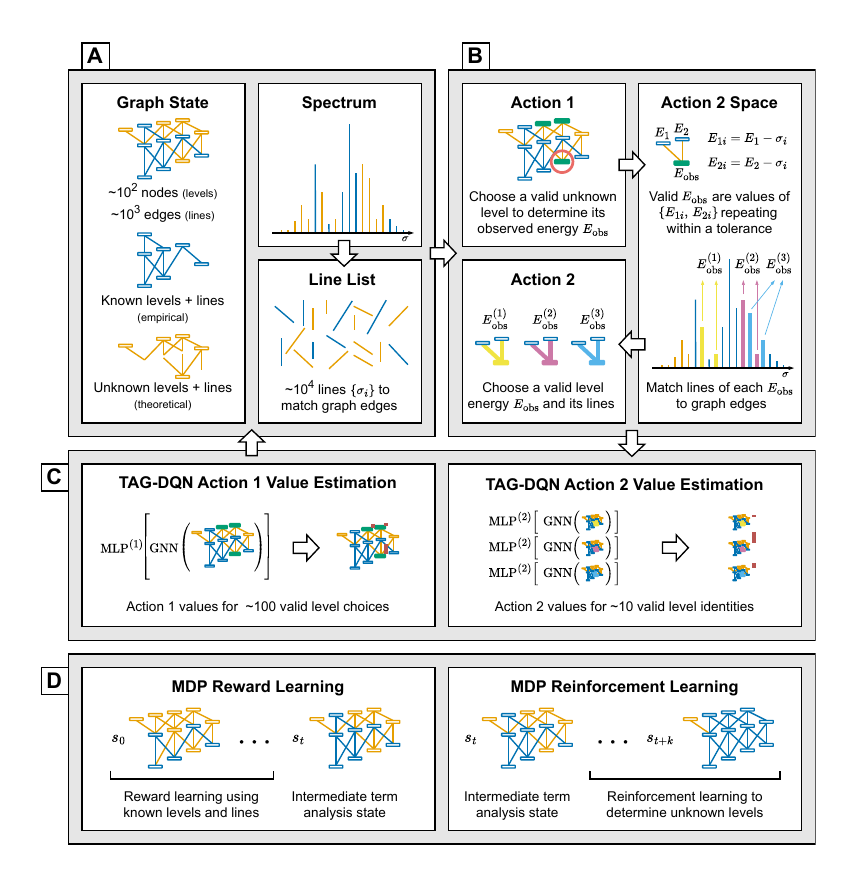}
    \caption{\textbf{Illustration of the MDP environment and \algshort.} \textbf{A} The term analysis state is represented as a graph with node and edge features, alongside the spectral line list. \textbf{B} Actions alternate between two regimes; each pair of actions leads to the determination of the observed energy $E_\text{obs}$ for one level by matching at least two lines from the line list to unknown edges in the graph. \textbf{C} \algshort~employs a GNN to embed graph representations, which are inputs for multilayer perceptrons (MLPs) estimating Q-values for each action (vertical bars); the highest Q-value action advances the MDP to the next state. \textbf{D} Given a term analysis state $s_t$, the MDP trajectory leading to $s_t$ involving known levels is used for reward learning, while RL with the learned reward function guides the discovery of unknown levels in future states $s_{t+k}$.}
    \label{fig1}
\end{figure}

We present our formulation of term analysis as an MDP for RL and evaluate TAG-DQN performance in realistic scenarios. Term analysis is typically carried out for one atomic species at a time, involving data on its known (empirically determined) and unknown (theoretically predicted) levels and lines, and an experimental spectral line list containing possible energy differences between any two levels. An overview of our methods is illustrated in Fig.~\ref{fig1}, which will be referenced throughout this paper. 

\subsection{MDP Definition}
An MDP is defined as the tuple $\langle \mathcal{S}, \mathcal{A}, R, \mathcal{T}, 
\gamma \rangle$ comprising the state space, action space, reward function, transition function, and discount factor~\cite{sutton2018reinforcement}. At a discrete time step $t$, the agent finds itself in state $s_t \in \mathcal{S}$, takes an action $a_t \in \mathcal{A}$, receives a reward $r_t \sim R(s_t,a_t)$, and advances to the next state $s_{t+1}$ determined by the transition function $\mathcal{T}(s_t,a_t, s_{t+1})$. The goal is to maximise the expected cumulative discounted reward 
\begin{equation}
     \mathbb{E}[G_t]=\mathbb{E}\Bigg[\sum_{k=0}^H\gamma^k r_{t+k}\Bigg],
 \end{equation}
where $\gamma \in [0, 1]$ is the discount factor trading off immediate and future rewards. The policy $\pi(a|s)$ is a probability distribution of actions given states determining the agent's behaviour. We also define the state-action value function of a policy $\pi$ as $Q_\pi(s_t,a_t)=\mathbb{E}_\pi[G_t|s_t,a_t]$; values for particular states and actions are referred to as Q-values. Our MDP is episodic, and $t$ is limited to a finite horizon $H$ for computational feasibility, which is set manually to specify the maximum number of unknown energy levels that can be determined within a single episode, starting from the initial MDP state.

\subsection{State Space $\mathcal{S}$} \label{sec:state}
The state is defined by the data incorporated in a term analysis, as illustrated in Fig.~\ref{fig1}A. Known spectral lines, known energy levels, and theoretical calculations form a graph $\mathcal{G} = \langle \mathcal{V}, \mathcal{E} \rangle$ where nodes $v \in \mathcal{V}$ represent levels and edges $e \in \mathcal{E}$ represent lines. Nodes relate to edges through $\sigma = E_u - E_l$, with the ground level energy fixed at zero. The goal of term analysis is to expand the subgraph of known levels by replacing unknown parts of the graph using data from observed spectra, which are typically preprocessed into a line list by fitting $\sim10^4$ model spectral line profiles. Each line list entry corresponds to an electric dipole fine structure transition, mainly characterised by observed wavenumber $\sigma_{\text{obs}}$, standard wavenumber uncertainty $\delta\sigma_{\text{obs}}$, relative intensity $I_{\text{obs}}$, and signal-to-noise ratio $S/N_{\text{obs}}$. These quantities remain unaltered by the MDP and serve to generate actions. 

The graph nodes and edges are described by features. A known line is a theoretical atomic transition (edge) matched to an entry in the line list, and a known level has energy determined via weighted least-squares optimisation of observed known line wavenumbers. The subgraph of known levels is typically connected, so all observed energies $E_{\text{obs}}$ are relative to the zero energy ground level. The node features are 
\begin{equation}\label{eq:nodefeats}
    \mathbf{x}_v = [E_{\text{calc}}, \:E_{\text{obs}}, \:\text{known}, \:\text{selected}],
\end{equation}
with theoretical energy as $E_{\text{calc}}$ and binary flags for ``known" and ``selected" levels. Unknown $E_{\text{obs}}$ values are zero. The ``selected" feature indicates the level currently being determined. The edge features are
\begin{equation}\label{eq:edgefeats}
\mathbf{x}_e = [\sigma_{\text{calc}}, \:\sigma_{\text{obs}}, \:\delta\sigma_{\text{obs}}, \:I_{\text{calc}}, \:I_{\text{obs}}, \:gA_{\text{calc}}, \:S/N_{\text{calc}}, \:S/N_{\text{obs}}, \:\rho, \:\text{known}],
\end{equation}
representing theoretical and observed wavenumber, observed wavenumber uncertainty, theoretical and observed relative intensities on the same scale, theoretical weighted transition probability, expected and observed $S/N$, observed local line density $\rho$ (number of lines per cm$^{-1}$), and a binary ``known" flag. Observed quantities are zero if ``unknown" and Boltzmann level populations are assumed for $I_{\text{calc}}$ estimation. Further details on graph feature design are provided in Methods.

\subsection{Action Space $\mathcal{A}$} \label{sec:action}
As illustrated by Fig.~\ref{fig1}B, the MDP alternates between two types of actions $\langle a^{(1)}, a^{(2)} \rangle$ in order to maintain feasible MDP branching factors. 
The pair of actions results in the determination of $E_{\text{obs}}$ for one unknown level and matching at least two observed spectral lines to corresponding edges connecting it to known levels.

The first action type $a^{(1)}$ selects an unknown level to determine and flips the ``selected" flag of the corresponding node. The action space $\mathcal{A}^{(1)}$ contains only unknown levels with at least two connections to known levels. While level energy determination by a single line is possible for a few levels in human analyses, we exclude these to avoid unfeasibly large $\mathcal{A}^{(1)}$ as such cases are highly situational. If $|\mathcal{A}^{(1)}|=0$, the MDP terminates. In our case studies, $|\mathcal{A}^{(1)}|$ ranges between 0 and 200 depending on MDP environment parameters.

The second action type $a^{(2)}$ matches at least two lines from the line list to edges between the unknown level selected by $a^{(1)}$ and known levels. To determine the action space $\mathcal{A}^{(2)}$ (possible $E_{\text{obs}}$ values), we first gather the $k>1$ graph edges connecting the selected level to $k$ known levels ($k=2$ for Fig.~\ref{fig1}B). For each edge, candidate matches in the line list are filtered by 
\begin{equation} \label{eq:linelist_tol}
    \sigma_{\text{obs}}\in[\sigma_{\text{calc}}\pm\mathit{\Delta}E]
    \quad \text{and} \quad
    I_{\text{obs}}\in[I_{\text{calc}}\pm\mathit{\Delta}I],
\end{equation}
where $\mathit{\Delta}E$ and $\mathit{\Delta}I$ are theoretical uncertainties for level energy and line intensity, respectively. The filtered $\sigma_{\text{obs}}$ values of an edge are added to, or subtracted from, the known level energy of that edge, depending on whether the known level is predicted as the lower or upper level. This generates candidate $E_{\text{obs}}$ values from each edge. Since $k>1$, $E_{\text{obs}}$ candidates not repeating within a tolerance $\delta E$ (approximately the maximum experimental wavenumber uncertainty) are neglected. The number of candidates can be estimated by
\begin{equation} \label{eq:A2}
N_{\text{cand}}(k) \sim \frac{\mathit{\Delta}E}{\delta E} \Bigg(\prod_{i=1}^k(1+2\,\delta\mkern-1mu E\,\rho_i) - 1 - \sum_{i=1}^k 2\,\delta\mkern-1mu E\,\rho_i\Bigg),
\end{equation}
where $\rho_i$ is estimated as the number of filtered line list entries divided by $2\,\mathit{\Delta} E$. For typical values ($\mathit{\Delta}E=500$~cm$^{-1}$, $\delta E=0.05$~cm$^{-1}$, $\rho=0.1$ per cm$^{-1}$), $N_{\text{cand}}(k=2)\sim5$. The base $|\mathcal{A}^{(2)}|$ is $N_{\text{cand}}(k)$ and can reach $10^3$ for larger $k$, $\mathit{\Delta}E$, $\delta E$, and $\rho$. In practice, increasing the minimum $E_{\text{obs}}$ repetition count $N_{\text{rp}}$ reduces the action space when more than two spectral lines involving connecting known levels are expected significantly above the spectrum noise level. We raised $N_{\text{rp}}$ to 3 or 4 if more than 3 or 4 of the $k$ lines exceed $S/N_{\text{calc}}=5$, effectively neglecting terms with products of 2 or 3 $\rho_i$ in (\ref{eq:A2}). In our case studies, the median $|\mathcal{A}^{(2)}|$ is below 10. 

\subsection{Transitions $\mathcal{T}$} \label{sec:transitions}
The graph structure remains fixed throughout the MDP, but node and edge features are updated by actions, as illustrated by Fig.~\ref{fig1}. Transitions are deterministic; a single future state $s'$ can be reached with probability $1$ after taking action $a$ in state $s$. Concretely, each next state $s_{t+1}$ is generated by updating the ``known" and ``selected" features of the level chosen by $a^{(1)}$, re-optimising $E_{\text{obs}}$ for all known levels, and updating $[\sigma_{\text{obs}}, \delta\sigma_{\text{obs}}, I_{\text{obs}}, S/N_{\text{obs}}, \text{known}]$ of the new known edges using corresponding entries in the line list, which are excluded from $\mathcal{A}^{(2)}$ for the remainder of the episode. A no-op action for $a^{(2)}$ is always available, handling levels with no candidates and the forfeit of $a^{(1)}$. When no-op is chosen, the next state reverts to the state prior to $a^{(1)}$, but the time step increments and the level chosen by $a^{(1)}$ is excluded from $\mathcal{A}^{(1)}$ until the episode end. 

\subsection{Reward Function $R$} \label{sec:reward}
As term analysis is inherently empirical, the reward reflects confidence in level energies, which often have ambiguous values. This confidence arises from the ability of the determined level to enable future level determinations and its agreement with theory and observations. The consequences of choosing a level energy will be explored by RL. While the action space incorporates theoretical and experimental uncertainties, a non-trivial reward remains necessary to differentiate between term analysis states. As $a^{(1)}$ and $a^{(2)}$ have different semantics, we also differentiate their reward functions. 

Term analysis prioritises lines with the highest signal-to-noise ratios (equivalently, minimising uncertainties and entropy~\cite{sivia2006data}),  constraining subsequent analysis of weaker lines. Thus for $a^{(1)}$, we define reward $r^{(1)}$ as
\begin{equation} \label{eq:R1}
    r^{(1)} = 0.1 \cdot \log_{10} \Bigg(\sum^{k}_{i=1} S/N_{\text{calc}, i}\Bigg),
\end{equation}
where $S/N_{\text{calc},i}\in[2,10^4]$, the sum is over the $k$ edges between the selected unknown level and known levels, and the 0.1 factor scales reward magnitude for learning stability. For $a^{(2)}$, the reward is
\begin{equation} \label{eq:R2}
    r^{(2)} = (D-1) \cdot r^{(1)},
\end{equation}
with $D \in [0,1]$ representing preference score for a particular $a^{(2)}$. The no-op $a^{(2)}$ has $D=0$, yielding net zero reward; otherwise, the net reward is positive. 

Defining $D$ is challenging. In practice, human experts weigh the agreement between observed and theoretical line intensities and the consistency of repeated $E_{\text{obs}}$ values within wavenumber uncertainties. However, uncertainties in transition probabilities and errors in spectral line fitting, especially for widespread weak and/or blended lines, complicate this judgement. While these nuances can be verified by spectrum inspection or improving theoretical calculations, they can also be inferred from properties of the $k$ lines from experience. 

Thus, we learn $D$ from historical human expert decisions via a form of inverse reinforcement learning~\cite{ng_algorithms_2000}, approximating part of the human reward function. We use a simple multi-layer perceptron (MLP) to predict $D$ from features of the $k$ lines
\begin{equation}
    [\delta\sigma_{\text{obs}}, \:\sigma_{\text{diff}}, \:\delta\sigma_{\text{obs}} - \sigma_{\text{diff}}, \:I_{\text{obs}}, \:I_{\text{calc}}, \:-|I_{\text{obs}} - I_{\text{calc}}|, \:\rho],
\end{equation}
where $\sigma_{\text{diff}}<\delta E$ is the smallest wavenumber difference between $E_{\text{obs}}$ of this line and $E_{\text{obs}}$ of the other $k-1$ lines. To train the model, we generate expert MDP state transitions $\langle s_t,a_t,r_t,s_{t+1}\rangle$ for supervised learning by stochastically ``reversing" the MDP from an intermediate term analysis state (Fig.~\ref{fig1}D; see Methods for details).

To evaluate the reward function independently of RL, we considered a normalised ranking metric to account for variable $|\mathcal{A}^{(2)}|$. The \textit{fractional rank} metric is defined as the rank of the expert action among the predicted $D$ scores, normalised by $|\mathcal{A}^{(2)}|$. A score of 1.0 indicates the expert action was top-ranked and random ranking yields $0.5$ in expectation. For our validation dataset, the fractional rank averaged $0.91$ under a median $|\mathcal{A}^{(2)}|=87$ ($N_{\text{rp}}$ fixed at 2). Thus, expert actions were frequently ranked among the top predicted $D$ scores. 

\subsection{Case Study MDP Environments and Key Parameters}
We investigated four MDP environments from Co~II, Nd~III, and Nd~II term analyses using FT spectral line lists from published studies \cite{pickering1998, ding2024}. We used atomic structure and spectrum calculations that are only achievable given the known levels of the initial state of each environment to accurately represent the challenges of term analysis. Reducing MDP complexity is key for feasibility and RL performance. We removed graph edges with $S/N_{\text{calc}}<2$, excluded line list entries already matched to edges from $\mathcal{A}^{(2)}$, fixed initial state known level energies in level energy optimisations, limited spectral ranges to target groups of levels, and neglected levels unlikely to be determinable (e.g., far configurations). A maximum cap on $|\mathcal{A}^{(2)}|$ is set for efficient exploration and memory control. If an $a^{(1)}$ induces an $|\mathcal{A}^{(2)}|$ exceeding this cap, a no-op is enforced. 

\begin{table}[t]
    \centering
    \caption{Key MDP environment parameters.}
    \vspace{0.5em}
    \resizebox{\textwidth}{!}{
    \begin{tabular}{lcccccccccc}
         \hline
          Case & Range$^{a}$ & $N_{\text{lin}}$$^{a}$ & $(|\mathcal{V}|, |\mathcal{E}|)$$^{b}$ & $H$& $\mathit{\Delta}E$ & $\mathit{\Delta}I$& $\delta E$  & $|\mathcal{A}^{(1)}|$ & $|\mathcal{A}^{(2)}|$ & $|\mathcal{A}^{(2)}|$\\
          & ($10^3$~cm$^{-1}$) & ($10^3$) & & & (cm$^{-1}$) & & (cm$^{-1}$) & range$^{c}$ & median$^{c}$ & max\\
         \hline
         Nd III & $30-55$  &  3 & (600, 1000) & 128 & 1500 & 1.0 & 0.05 & $10-60$  & 6 &  256\\
         Co II  & $34-83$  &  4 & (900, 4000) & 128 & 1500 & 1.2 & 0.05 & $90-150$ & 3 &  256\\
         Nd II u& $10-55$  & 22 & (700, 1800) &  64 & 3000 & 1.2 & 0.05 & $60-200$ & 5 &   32\\
         Nd II k& $10-55$  & 22 & (500, 2200) & 512 &  250 & 1.0 & 0.05 & $0-200$  & 3 &   32\\
         \hline
    \end{tabular}
    }
    \begin{flushleft}
        \small
        $^{a}$ Line list spectral range, within which $N_{\text{lin}}$ lines were used for $\mathcal{A}^{(2)}$ determination. \\
        $^{b}$ Graph size in terms of (number of nodes, number of edges). \\
        $^{c}$ Measured over the history of \algshort~training, not from uniform random MDP state transitions.

    \end{flushleft}
    \label{tab:env_params}
\end{table}
Environment parameters are summarised in Table~\ref{tab:env_params}. For Nd III, the initial state includes observed FT spectral lines and energies of 40 4f$^3$(4f, 5d) levels revised from \cite{ryab2006}, excluding 19,403~cm$^{-1}$ with $J=3$ as it was concluded erroneous \cite{ding2024}, with Cowan code \cite{cowan1981, kramida2021} calculations (Hartree-Fock method with relativistic corrections) parameterised using these 40 levels \cite{ryab2006}. For Co II, the initial state is the revision \cite{pickering1998} of 141 3d$^7$(3d, 4s, 4p, 5s, 4d) levels \cite{sugar1985}. We use Cowan code calculations parameterised using these levels \cite{priv_com}. 
For both Nd II u and Nd II k, the initial state includes the six 4f$^4$($^5$I)6s $^6$I ground term levels, the 4f$^4$($^5$I)6s $^4$I$_{9/2}$ level, five 4f$^4$($^5$I)6p $^6$K levels ($J=9/2$ to $17/2$), and 11 transitions between them. These were chosen for their relatively high eigenvector purities and line intensities. Nd II u uses only \textit{ab initio} calculations \cite{gaigalas2019}, while Nd II k uses Cowan code calculations parameterised using all known levels, representing revision of known energies \cite{blaise1984} using more precise measurements. For efficiency, doubly-excited configurations of Nd~III, and Nd~II levels predicted above 35,000~cm$^{-1}$, were excluded from the graphs. Setting the $|\mathcal{A}^{(2)}|$ cap was also necessary for Nd~II due to large $N_{\text{lin}}$.

The episode length $H$ balances computational feasibility with the ability of the agent to learn beyond short-term consequences. Additionally, an upper limit on $H$ ensures meaningful analysis, as atomic structure calculations can be improved after determining several dozen levels. We found $H=128$ effective for Nd~III and Co~II, with no improvement in performance over $H=64$ for Nd~II~u, and $H=512$ viable for Nd~II~k due to lower $\mathit{\Delta} E$ (set by domain knowledge). The $\mathit{\Delta}I$ range is estimated at one order of magnitude to account for uncertainties from theoretical transition probabilities and Boltzmann level populations. $\mathit{\Delta}I$ is higher for Co~II due to averaging line lists and charge-transfer effects \cite{johansson1980}, and for Nd~II~u due to \textit{ab initio} calculations. The repeating candidate level energy tolerance $\delta E$ approximately matches the highest $\delta \sigma_{\text{obs}}$. 

\subsection{\algshort~Summary} \label{sec:agent}
\algshort~is a model-free, value-based method based on the DQN~\cite{mnih2015}, as illustrated in Fig.~\ref{fig1}C. The agent estimates Q-values of valid actions, from which a policy $\pi$ can be derived by selecting the $a_t$ with the largest Q-value. Experience tuples $\langle s, a, r, s' \rangle$ are stored in a \textit{replay buffer}. Model parameters are adjusted by stochastic gradient descent using losses from experience batches sampled from the buffer, with respect to a periodically updated \textit{target network}. We chose this algorithm class for its higher sample efficiency compared to policy gradient approaches. Furthermore, as we only aim to find a maximum-reward trajectory, a ``greedy'' policy $\pi$ with respect to the estimated state-action values is sufficient. Large state and action spaces require deep neural networks for function approximation; we opt for GNNs as a learning representation, as successfully leveraged by other recent works treating graph combinatorial optimisation problems with RL~\cite{darvariu2024grl}. We also adopt several widely accepted DQN extensions~\cite{hessel_rainbow_2018} including duelling~\cite{wang2016}, double Q-learning~\cite{hasselt2016}, multi-step returns~\cite{sutton1988}, and noisy networks for exploration \cite{fortunato2018}. Implementation and hyperparameter details are in Methods.

\subsection{Level Determination Accuracy and Baseline Methods}
Several solution methods apply to the term analysis MDP. We also evaluate two baseline agents: a \textit{greedy search} agent which always chooses the maximum reward action, and a standard \textit{Monte-Carlo tree search} (MCTS) agent \cite{browne2012} with exploration by upper confidence bound for trees (UCT) \cite{kocsis2006}. Due to large state and action spaces, these discrete search methods operate with a limited horizon and may choose myopic actions. We evaluate \algshort~and MCTS across $25$ different random seeds and report average performance, while greedy search is deterministic. Unless otherwise stated, we report 95\% confidence intervals in all tables and figures. MCTS hyperparameters are tuned as detailed in Methods. Performance is assessed using the following metrics:

\begin{itemize}
    \item $R_\text{max}$: Maximum cumulative reward acquired in a MDP episode with length $H$.
    \item $N_{\text{c}}$: Number of levels determined in the the $R_{\text{max}}$ episode whose $E_{\text{obs}}$ match literature values within $\delta E$.
    \item Accuracy (Acc.): Ratio between $N_{\text{c}}$ and the total number of levels determined in the $R_{\text{max}}$ episode, which is $N_{\text{c}}/\text{Acc.}\leq\frac{H}{2}$. A ratio of 1 implies $E_{\text{obs}}$ agreement with published values for all levels determined in the $R_\text{max}$ episode.
\end{itemize}

\begin{table}[t]
    \centering
    \caption{Results and comparisons with benchmark agents. \algshort~matches MCTS performance as judged by downstream metrics in 2/5 cases and outperforms it in 3/5 cases, while Greedy search is worse overall.}
    \vspace{0.5em}
    \resizebox{\textwidth}{!}{
    \begin{tabular}{l|ccc|ccc|ccc}
         \hline
          Case  & \multicolumn{3}{c|}{Greedy search} & \multicolumn{3}{c|}{MCTS}   & \multicolumn{3}{c}{\algshort~}   \\
                &  $R_\text{max}$ & $N_{\text{c}}$  & Acc.  &  $R_\text{max}$ & $N_{\text{c}}$  & Acc.  &  $R_\text{max}$ & $N_{\text{c}}$  & Acc.  \\
         \hline                                                                    
         Nd III &  7.3 &  28    & 0.52 &  8.5\tiny{$\pm 0.9$}     &   37\tiny{$\pm 4$} & 0.58\tiny{$\pm 0.06$}    & 7.2\tiny{$\pm 0.3$}   &37\tiny{$\pm 5$}   & 0.59\tiny{$\pm 0.08$}  \\
         Co II  & 9.9 &  41    & 0.84 & 14.9\tiny{$\pm0.7$}     &   61\tiny{$\pm 2$} & 0.97\tiny{$\pm 0.03$}    &13.2\tiny{$\pm 1.1$}   &60\tiny{$\pm 3$}   & 0.95\tiny{$\pm 0.02$}  \\
         Nd II u&  6.6 &   8    & 0.38 &  8.4\tiny{$\pm 0.1$}     &    9\tiny{$\pm 6$} & 0.30\tiny{$\pm 0.18$}    & 6.8\tiny{$\pm 0.7$}   &15\tiny{$\pm 6$}   & 0.54\tiny{$\pm 0.21$}  \\
         Nd II k& 51.3 & 184    & 0.79 & 53.7\tiny{$\pm 2.1$}    & 185\tiny{$\pm 9$} & 0.77\tiny{$\pm 0.02$}    &48.8\tiny{$\pm 0.8$}  &210\tiny{$\pm 7$} & 0.87\tiny{$\pm 0.03$}  \\
         Nd II k corr. label&  - & 132& 0.57 &     -       & 130\tiny{$\pm 11$} & 0.54\tiny{$\pm 0.04$}    & -        &169\tiny{$\pm 10$} & 0.69\tiny{$\pm 0.05$}  \\
         \hline 
    \end{tabular}
    }
    \label{tab:results}
\end{table}
Main results for the four case studies are in Table~\ref{tab:results}. Generally, the highest reward episode corresponded to the highest $N_{\text{c}}$. For Nd II k, more accurate calculations for known levels allowed a reliable secondary metric involving level labels such as $J$, term symbol, and eigenvector composition, distinguished by level index in the calculations. This is given in the final row of Table~\ref{tab:results}, where $N_\text{c}$ is instead the number of determined levels with $E_{\text{obs}}$ agreeing with accepted values and human assigned level indices. Level determination accuracy (Acc.) is concluded to be dependent on the accuracy of theoretical calculations and the alignment between reward and $N_{\text{c}}$. Performance was best in the Co~II environment, as expected given its simpler atomic structure and strong influence on reward learning (see Methods). 

Greedy search consistently underperformed compared to RL agents. Notably, MCTS achieved higher $R_{\text{max}}$ than \algshort, yet \algshort~reached higher upper bounds of $N_{\text{c}}$, significantly in the Nd II cases. We interpret this as the ability of \algshort~to choose level identities that are more likely to remain consistent with observations and theory throughout the episode, whereas MCTS rollouts were shallow and favoured short-term rewards, yielding higher trajectory rewards. This is supported by the significantly larger number of correctly labelled levels for \algshort~in Nd II k. Incorrect initial level labels are also common for humans, though likely less frequent. 

\begin{figure}
    \centering
    \includegraphics[width=0.8\linewidth]{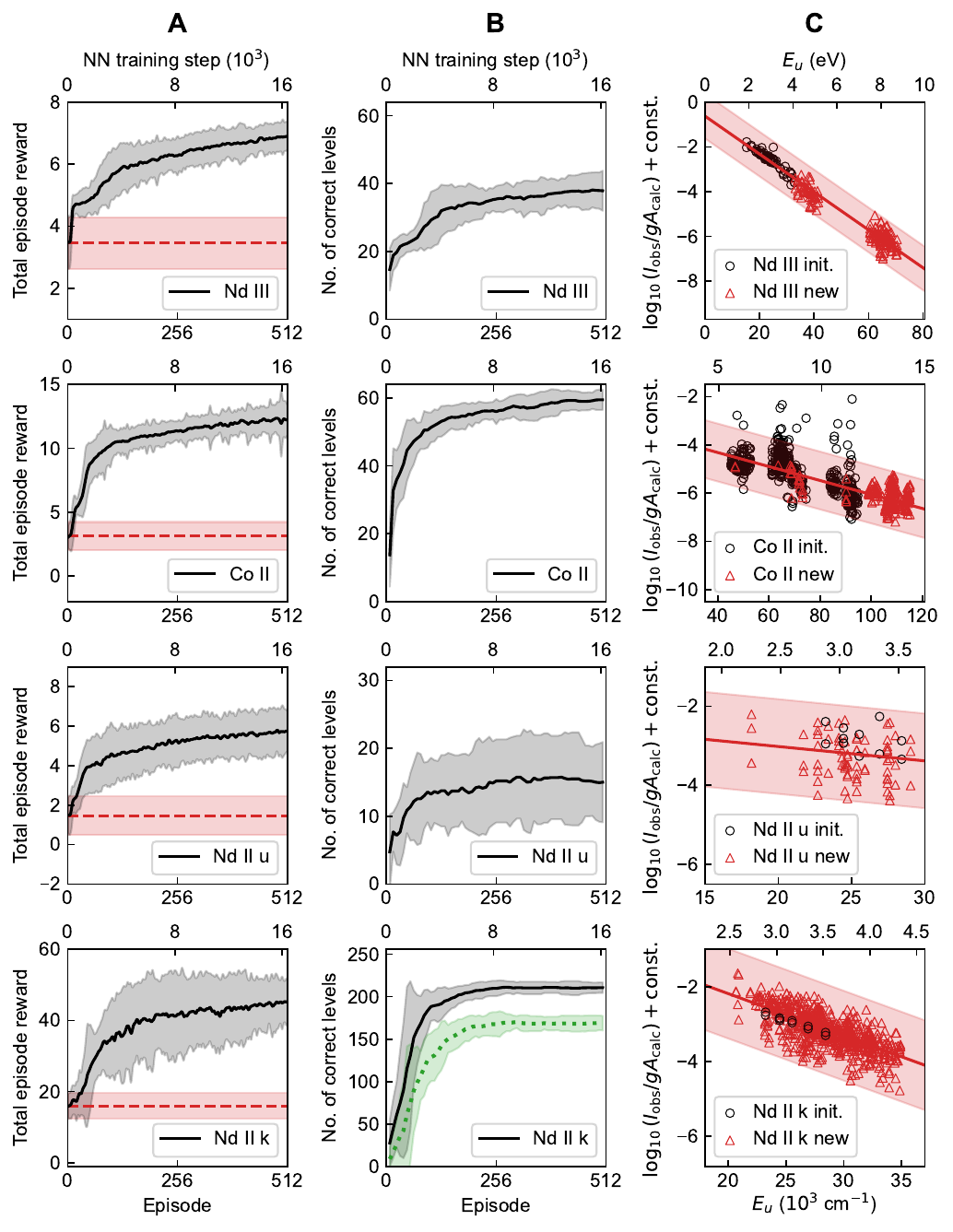}
    \caption{\textbf{\algshort~results for the four case studies.} \textbf{A} Learning curves (i.e., cumulative rewards obtained as a function of training steps). The dashed horizontal lines show reward obtained during the pre-training episodes and correspond to rewards obtained by choosing actions in the MDP uniformly at random. \textbf{B} $N_{\text{c}}$ of the final state of the most recent maximum reward episode. The dotted curve for Nd II k shows the number of correct levels that also match human chosen level labels, and the y-axes limits are $\frac{H}{2}$. \textbf{C} Boltzmann plot (log of relative level population against upper level energy $E_u$) using known lines from the final state of the maximum reward episode of a single seed; circles show initial state known lines, triangles show new known lines from RL, and the shaded region of the linear fit shows $\pm\mathit{\Delta} I$.}

    \label{fig:results}
\end{figure}

Learning curves and analyses for \algshort~in each case study are shown in Fig.~\ref{fig:results}. 
From Fig.~\ref{fig:results}A and B, reward alignment with $N_{\text{c}}$ is evident. Levels of different electron configurations were also determined, as shown in Fig.~\ref{fig:results}C. The Boltzmann level population and $\mathit{\Delta} I \sim 1$ provides a reasonable basis for line intensity filtering for $\mathcal{A}^{(2)}$, even for Co II where line intensities were averaged over two plasma sources, one exhibiting charge-transfer population enhancements for 4d levels \cite{pickering1998, johansson1980}. However, lines with observed intensities deviating significantly from the Boltzmann assumption were excluded from $\mathcal{A}^{(2)}$, as shown in Fig.~\ref{fig:results}C.

\subsection{Ablation Studies}
Standard DQN extensions for \algshort~were individually disabled in the Nd~III environment to assess their impacts.
\begin{figure}
    \centering
    \includegraphics[width=.6\linewidth]{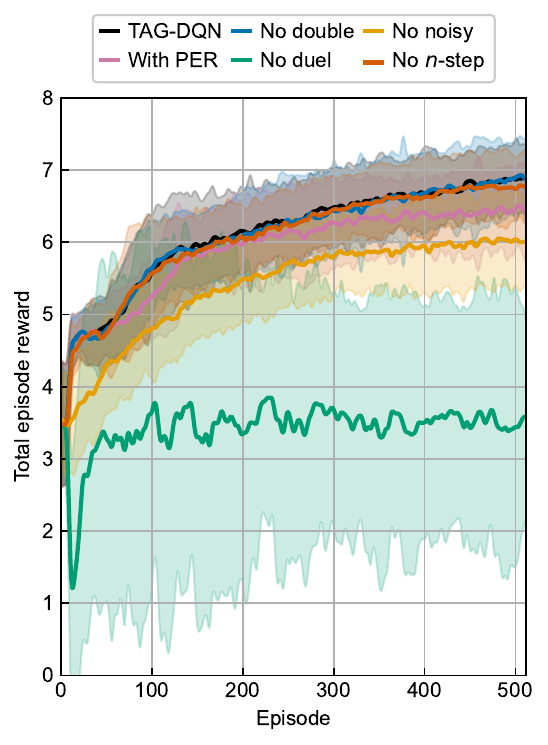}
    \caption{\textbf{Learning curves of \algshort~(black) and its variants in the Nd~III environment.} Duelling is key to good performance while PER was actively harmful.}
    \label{fig:ablation}
\end{figure}
Figure~\ref{fig:ablation} shows that all extensions except double Q-learning and prioritised experience replay (PER) improved performance. Double Q-learning was inconclusive despite being recommended in the RL literature, and PER was found to be harmful across all case studies.
We attribute this to the smaller buffer size (due to memory limits of storing large graphs) and fewer time steps for convergence compared to environments where PER was shown beneficial~\cite{schaul2016}. Duelling is critical for \algshort, likely because of the large action spaces, often hundreds in size per step.
\section{Discussion}

As the aim of term analysis is to maximise the number of correctly determined level energies $N_{\text{c}}$, deploying \algshort~would involve training reward functions for the specific MDP and using results from the highest $R_{\text{max}}$ run. Tuning hyperparameters (if resources permit) may further improve performance given those tuned for the Nd~III environment underperformed in Nd~II k, which differs substantially in $H$ and $\mathit{\Delta} E$. As Nd~II k is the first revision of Nd~II level energies using FT spectra, we include the new levels and lines determined by \algshort~in the online repository. However, we emphasise that this Nd~II data is tentative, as acceptable term analysis requires spectrum line profile inspections and improved semi-empirical calculations for validation, which are in progress in one of our current projects. 

By design, the final MDP state after $H$ time steps becomes a new initial state for determining additional levels, though human intervention is recommended. This stage allows spectrum inspection, exclusion of poorly fitted lines from level energy optimisation, pruning incorrect levels, refining semi-empirical calculations and level labels, and retraining the reward function. As our reward function is trained only using data from the Co II MDP (see Methods), a robust reward function trained using a large variety of line lists and MDPs would be a logical improvement.

\begin{figure}
    \centering
    \includegraphics[width=.6\linewidth]{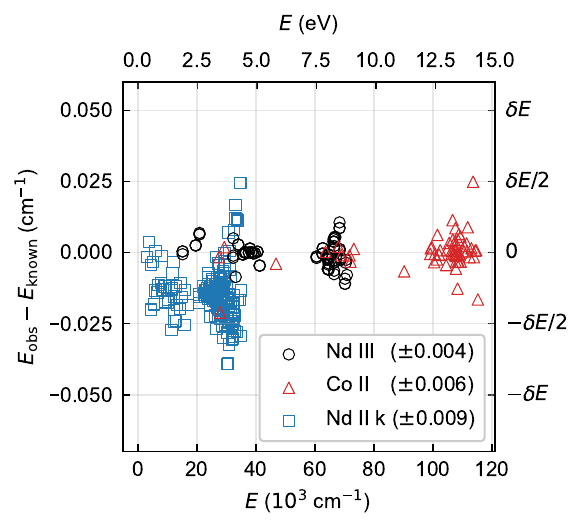}
    \caption{\textbf{Difference between determined level energies and their known (accepted) values for a single seed for each species}. Only levels contributing to $N_{\text{c}}$ are shown. Level energy differences for Nd~III, Co~II, and Nd II k are shown in circles, triangles, and squares, respectively. The root-mean-square energy differences are given in parentheses of the legend. Typical level energy uncertainty by FT spectroscopy is of order 0.001~cm$^{-1}$. The offset of Nd II k energies are expected and within uncertainties as their known values were derived by lower precision grating spectroscopy \cite{blaise1984}.}
    \label{fig:dE}
\end{figure}

We used $\delta E$ to determine $\mathcal{A}^{(2)}$ and the reward function to address influence from spectral lines with outlier wavenumbers (e.g., unknown line blending or poor spectral line fitting). However, shifts to determined level energies from such outliers are not negated, as shown in Figure~\ref{fig:dE},
where a small fraction of level energies significantly deviate from their accepted values but remain within $\delta E$. A priority improvement could incorporate the spectrum or a dynamic line list in the MDP, so that $\mathcal{A}^{(2)}$ determinations could involve automated line profile analysis (e.g., \cite{ding2024}), which could also reduce the need human validation. We note that these level energy deviations are negligible for most applications at spectral resolving powers $\ll10^6$.

Our case studies and methods span diverse term analysis scenarios but omit the most challenging situations, such as single-line level determinations or joining disconnected known-level subgraphs, including term analyses starting with no known levels (applicable to very few low-ionisation species). In these cases, MDP complexity increases significantly, likely requiring careful MDP redesign, where action space reductions from spectra of various plasma conditions or instruments could also be included. Our approach is extendable to other experimental methods (e.g., grating spectroscopy) with minor environment adjustments such as modifying $\delta E$, $\mathit{\Delta} I$, and the reward function. Lastly, evaluating against published, known levels was far more feasible in our scope compared to carrying out several new term analyses for unknown levels for evaluation. Thus, we anticipate future applications of our methods to better demonstrate their true potential.

Automated fine structure level energy determination is now possible, assisting experts in analyses requiring months to years of effort with traditional techniques. Despite the reduced accuracy compared to humans and dependence on atomic structure calculations, we expect our methods to become a cornerstone in addressing the ever-increasing demand for atomic data. Beyond atomic spectroscopy, the present work is testament to the potential of graph reinforcement learning~\cite{darvariu2024grl} and AI in general~\cite{wang2023scientific} to assist scientific discovery, particularly when co-developed by scientists and AI researchers. We hope that this work will encourage AI researchers to consider scientific problems as valuable testbeds for developing AI techniques, while also showing scientists how RL can potentially be harnessed for their data- and labour-intensive tasks.



\section{Methods}

\subsection{Level Energy Optimisation}
Observed energies $E_{\text{obs}}$ of $N$ known levels are determined from $M$ known lines by weighted least-squares minimisation
\begin{equation}\label{eq:lopt}
E_{\text{obs}}=\underset{E_n}{\mathrm{argmin}}\Bigg[\sum_{m=0}^{M-1} w_m\bigg(\sum^{N-1}_{n=0}S_{mn} E_n - \sigma_m\bigg)^2\Bigg],
\end{equation}
where $w_m=(\delta\sigma_m)^{-2}$, $\sigma_m$ and $\delta\sigma_m$ are the $\sigma_{\text{obs}}$ and $\delta\sigma_{\text{obs}}$ of line $m$, and $S_{mn}$ defines the relationship
\begin{equation}
    \sum^{N-1}_{n=0}S_{mn} E_n - \sigma_m = E_{u} - E_{l} - \sigma_m,
\end{equation}
with $E_{u}$ and $E_{l}$ as the upper and lower level energies of line $m$. The ground level energy $E_0$ is fixed at zero.

\subsection{Graph Feature Design}
Level parity and total angular momentum $J$ are distinguished by the unique node indices but excluded from node features, as their selection rules \cite{cowan1981} are already encoded by the graph structure, and transition probabilities are edge features. Configuration labels, term labels, and eigenvector compositions are also omitted, since they change with improved semi-empirical atomic structure calculations and are less meaningful under high level densities. 

Upper level Boltzmann populations fitted using known lines are extrapolated to estimate $I_{\text{calc}}$ on the same scale as $I_{\text{obs}}$. The $S/N_{\text{calc}}$ is derived from $I_{\text{calc}}$ and the estimated ratio between $S/N_{\text{obs}}$ and $I_{\text{obs}}$ as a function of wavenumber. The line density $\rho$ is the number of lines in the line list within $\sigma_{\text{calc}} \pm \mathit{\Delta}E$ with $S/N_{\text{obs}}$ within one order of magnitude of $S/N_{\text{calc}}$, divided by $2\,\mathit{\Delta}E$. To avoid underestimation for weak lines, $S/N_{\text{calc}}$ is clipped to 10 when computing $\rho$. All features relating to $\rho$, $S/N$, and $gA_{\text{calc}}$ are on the log-scale for stable learning.

\subsection{Reward Learning}
The MLP model for predicting $D$ embeds features of each new known line, aggregates the embeddings by summation to reflect higher confidence with more lines, and outputs $D\in[0,1]$ via a sigmoid function. Explicit feature engineering including feature differences improved generalisation to unseen states, compensating for MLP simplicity and limited training data. The intensities $I_{\text{obs}}$ and $I_{\text{calc}}$ are normalised by their respective maximum values within each set of $k$ lines, with 10\% noise added to $I_{\text{calc}}$ to account for calculations for known transitions being generally more accurate. Without intensity normalisation, the mean validation fractional rank was higher but led to lower $N_{\text{c}}$ during RL, likely because weaker lines were under-represented in the training dataset and unknown lines are typically weaker.

We collected 115 expert $a^{(2)}$ MDP state transitions from the Co~II environment as the training dataset and 23 from Nd~III as the validation dataset, by stochastically reversing the MDP from environment initial states. Each reverse step converts a known level and its lines to unknown, continuing until no cycles remain on the known-level subgraph. To invert the prioritisation of levels with the highest $S/N_\text{obs}$ lines, known level removal probabilities were assigned via a softmax over $-2r^{(1)}$, excluding levels whose removal would disconnect the known-level subgraph. This random reversal avoids reliance on unavailable human MDP trajectories and is valid since level determinations have no strict order. Levels outside cycles on the known-level subgraph are also removed before MDP reversal as they cannot appear in $\mathcal{A}^{(1)}$. After removal, the level is marked ``selected" in the prior state, with $\mathcal{A}^{(2)}$ computed at $N_{\text{rp}}=2$, which includes an expert action for reward learning. The next known level removal proceeds after reversing $a^{(1)}$ by flipping the ``selected" feature.

In each of the $115+23$ training samples, only the expert-chosen $a^{(2)}$ was labelled positive. The MLP model for $D$ contains 105 parameters, trained with Adam~\cite{kingma2015} to minimise weighted cross-entropy over 32 epochs. The model from the epoch with the lowest validation loss was selected and then used for RL in all four environments of Table~\ref{tab:env_params}. 

\subsection{TAG-DQN Learning Architecture}
The \algshort~agent consists of one GNN with parameters $\theta_1$ and four MLPs $(V_{\theta_2}, A_{\theta_3}, V_{\theta_4}, V_{\theta_5})$. We denote the union of all parameters by $\theta = \cup_i \{ \theta_i \}$ and the parameterised model as $Q_\theta$. Parameters $\theta$ are optimised by Adam~\cite{kingma2015} using a loss computed from a sampled experience batch once every $N_\text{SPT}$ steps. 

The graph state $s$, defined by nodes $\mathcal{V}$ and edges $\mathcal{E}$, is processed by multiple graph attention layers \cite{brody2022}, producing node embeddings $\text{GNN}(s)=\{\mathbf{h}_v\}_{v\in \mathcal{V}}$. Node and edge features $\mathbf{x}_v$ (\ref{eq:nodefeats}) and $\mathbf{x}_e$ (\ref{eq:edgefeats}), are used directly if continuous or one-hot encoded if discrete. An ELU activation \cite{clevert2016} follows each GNN layer except for the final one. Both $N_{\text{head}}$ and $H_{\text{GNN}}$ are the same for all layers, and outputs from each head are concatenated. 
The graph state embedding is obtained through mean aggregation 
\begin{equation}
    \mathbf{s}_{\text{agg}}=\frac{1}{|\mathcal{V}|}\sum_v\mathbf{h}_v.
\end{equation}
The embedding $\mathbf{s}_{\text{agg}}$ is concatenated with the normalised number of remaining episode steps and used as input for MLP value estimators. 

Separate state value MLP estimators $V_{\theta_{2}}$ and $V_{\theta_{5}}$ are used in duelling for $a^{(1)}$ and $a^{(2)}$ due to significantly different semantics of the two action types. For $a^{(1)}$, the node embedding of each valid level is concatenated with $\mathbf{s}_{\text{agg}}$ for representations $(s, a)$ for the advantage MLP estimator $A_{\theta_3}$. The Q-value for actions of type $a^{(1)}$ is estimated as
\begin{equation} \label{eq:duelling}
    Q(s, a) = V_{\theta_2}(\mathbf{s}_{\text{agg}}) + A_{\theta_3}(s, a) - \frac{1}{|\mathcal{A}^{(1)}|}\sum^{}_{a' \in \mathcal{A}^{(1)} }A_{\theta_3}(s, a').
\end{equation}
For $a^{(2)}$, we explicitly compute each next state $s'$, embed via the GNN, mean aggregate, and concatenate with normalised steps remaining to form $\mathbf{s}'_{\text{agg}}$ for advantage $A$ estimation
\begin{equation}
    A(s, a) = V_{\theta_4}(\mathbf{s}'_{\text{agg}}) - V_{\theta_5}(\mathbf{s}_{\text{agg}}),
\end{equation}
where $V_{\theta_4}$ is used to estimate the value of the next state $s'$, and the Q-value is estimated in duelling with the same formula as (\ref{eq:duelling}). This advantage estimation is possible as the transition function $\mathcal{T}$ is fully deterministic, allowing direct evaluation of $s'$ after applying $a^{(2)}$. 
All MLPs have two hidden layers of size $H_{\text{MLP}}$ with ReLU activations and a scalar output as $|\mathcal{A}^{(1)}|$ and $|\mathcal{A}^{(2)}|$ vary depending on $s$.

Parameters $\bar\theta$ of the target network $Q_{\bar\theta}$ are soft-updated after each update of the online network parameters $\theta$~\cite{lillicrap2016}
\begin{equation}
\bar\theta = (1-\tau) \bar\theta + \tau \theta    
\end{equation}
for $\tau \in[0, 1]$. Using $n$-step returns and double DQN, the loss from a replay buffer sample $\langle s_t,a_t,r_t,s_{t+n} \rangle$ is 
\begin{equation}\label{eq:loss}
 L=[\text{TD}^{[n]}]^2=[r_t^{[n]} + \gamma_t^{n}\displaystyle Q_{\bar\theta}(s_{t+n}, \underset{a_{t+n}}{\mathrm{argmax}}\,Q_{\theta}(s_{t+n},  a_{t+n})) - Q_{\theta}(s_t, a_t)]^2,    
\end{equation}
where
\begin{equation}
    r_t^{[n]}=\sum_{k=0}^{n-1}\gamma_t^{k}r_{t+k}.
\end{equation}

For exploration, all MLP output layers are noisy layers~\cite{fortunato2018}. Deterministic weights of noisy layers are initialised under a uniform distribution between $\pm (H_{\text{MLP}})^{-0.5}$, weights multiplying noise are initialised at $\sigma_0(H_{\text{MLP}})^{-0.5}$, where $\sigma_0\approx0.5$ is a hyperparameter. 

\subsection{Ablation Protocol}
To disable double Q-learning, the target network $Q_{\bar\theta}$ was used in (\ref{eq:loss}) for both action selection and value estimation~\cite{hasselt2016}. For ablating duelling, state value estimators $V_{\theta_2}$, $V_{\theta_5}$ and advantage calculations were removed, and $A_{\theta_3}$ and $V_{\theta_4}$ were used directly to estimate $Q(s,a)$. Noisy-network exploration was replaced by standard $\epsilon$-greedy exploration~\cite[Chapter~2]{sutton2018reinforcement} with $\epsilon$ decaying from 1 to 0.1 at a rate of 0.99 per episode. Setting $n=1$ disables $n$-step returns.

\subsection{Hyperparameter Tuning}

We tuned \algshort~hyperparameters in the Nd~III environment via grid search with the validation objective of maximising reward. Each set of hyperparameters was evaluated under 5 random seeds after training for 512 episodes ($\sim66$K steps). Table~\ref{tab:hyparams} summarises the search space and final settings. Due to high dimensionality, only a subset of all possible combinations indicated in Table~\ref{tab:hyparams} were investigated. For the Nd II k case study with much longer episodes and more accurate calculations, we instead used $n=1$ as significant improvements were observed over $n=2$.


\begin{table}[!h]
    \centering
    \caption{\algshort~parameter ranges and final hyperparameters.}
    \vspace{0.5em}
    \begin{tabular}{c|c|c}
         \hline
          Parameter & Search & Final \\
         \hline
          No. of GNN layers & [1, 2, 3, 4] & 3\\
          $H_{\text{GNN}}$  & [16, 32, 64] & 32 \\
          $N_{\text{head}}$ & [1, 2, 4, 8] & 4 \\
          $H_{\text{MLP}}$  & [8, 16, 32, 128, 256] & 32 \\
          Adam learning rate ($\times10^{-4}$) & [50, 10, 5, 1] & 10 \\
          Replay batch size & [4, 8, 16, 32] & 16 \\
          Soft target update rate $\tau$ & [0.05, 0.01, 0.005, 0.001] & 0.001 \\
          Steps per train  $N_\text{SPT}$ & [2, 4, 16] & $H/32$ \\
          Noisy nets $\sigma_0$ & [0.1, 0.5] & 0.5 \\
          Replay buffer capacity  $N_{\text{buffer}}$& 10K steps& 10K steps \\
          Min. history to start learning & 8 episodes& 8 episodes \\
          Multi-step returns $n$ &[1, 2, 3, 4] & 2\\
          Discount factor $\gamma$ &0.99 & 0.99\\
         \hline
    \end{tabular}
    \label{tab:hyparams}
\end{table}


The MCTS exploration and rollout depth parameters were optimised at 0.4 and 4 by grid search over the ranges [0.025, 0.1, 0.4, 0.8, 1.2] and [4, 8, 16], respectively, with 8 seeds per parameter pair. Random action sampling of vanilla MCTS is inefficient in large action spaces where only one action is correct, and low optimal rollout depth is expected as deeper rollouts introduce noise in the returns. For fair comparison between MCTS and DQN, the number of trials per timestep for MCTS is set at 512, and hence the two algorithms encounter the same number of MDP transitions.

\subsection{Implementation and Runtime Details}

The \algshort~agent is built using the \textit{PyTorch} \cite{paszke2019} and \textit{PyTorch Geometric} \cite{fey2019} libraries. Convergence of \algshort~is reached within 24 hours of training and inference time is negligible. For MCTS, we re-purpose an implementation originally designed for another graph combinatorial optimisation problem (causal structure discovery)~\cite{darvariu2025tree}. 

Experiments were carried out using the Imperial College high performance computing facility. Each task was allocated 8 CPU cores, up to 128 GB memory, and a 24 hour runtime. The full set of experiments presented in this paper, including hyperparameter tuning, would require about $10^5$ hours ($\approx11$ years) of hypothetical single-core CPU time.


\section*{Code and Data Availability}
The code and data used to generate results in this paper will be made available in a future version.


\bibliographystyle{unsrt}
\bibliography{bibliography}
\section*{Acknowledgements}
M.D. and J.C.P. acknowledge support from the Science and Technology Facilities Council (STFC) of the UK under grant numbers ST/N000939/1, ST/S000372/1, ST/W000989/1, and UKRI1188, and The Bequest of Prof. Edward Steers. V.-A.D. and N.H. acknowledge support from the Natural Environment Research Council (NERC) Twinning Capability for the Natural Environment (TWINE) Programme NE/Z503381/1, the Engineering and Physical Sciences Research Council (EPSRC) From Sensing to Collaboration Programme Grant EP/V000748/1, and the Innovate UK AutoInspect Grant 1004416. A.N.R is grateful to the late Dr J.-F. Wyart for help in Nd II calculations and to the support from research project FFUU-2025-0005 of the Institute of Spectroscopy of the Russian Academy of Sciences.

\section*{Author contributions}
M.D. -- planning, method design, execution, Co II calculations, manuscript preparation. \\V.-A.D. -- planning, method design, manuscript preparation. \\A.N.R. -- Nd II-III calculations and manuscript review. \\N.H. -- resource management, manuscript review. \\J.C.P. -- resource management, progress and manuscript review.


\end{document}